\documentclass{article}
\usepackage{spconf,amsmath,graphicx}
\usepackage{colortbl}
\usepackage{enumitem}
\usepackage[table,xcdraw]{xcolor}
\usepackage{graphicx}
\usepackage{multirow}
\usepackage[normalem]{ulem}
\useunder{\uline}{\ul}{}
\usepackage{multirow}

\usepackage{tikz}
\usepackage{comment}
\usepackage{amsmath,amssymb} 
\usepackage{bbm}
\usepackage{color}
\usepackage{xcolor}
\usepackage{enumitem}
\usepackage{booktabs}
\usepackage{multicol}
\usepackage{url}
\usepackage{mathtools, nccmath}
\newcommand*\rot{\rotatebox{90}}
\renewcommand{\vec}[1]{\mathbf{#1}}

\title{Text is All You Need:\\ Personalizing ASR Models using Controllable Speech Synthesis}

\name{Karren Yang*, Ting-Yao Hu*, Jen-Hao Rick Chang*, Hema Swetha Koppula*, Oncel Tuzel*
}
\address{*Apple}
\begin{document}
%
\setlength{\abovedisplayskip}{3pt}
\setlength{\belowdisplayskip}{3pt}

\maketitle
\begin{abstract}
Adapting generic speech recognition models to specific individuals is a challenging problem due to the scarcity of personalized data. Recent works have proposed boosting the amount of training data using personalized text-to-speech synthesis. Here, we ask two fundamental questions about this strategy: when is synthetic data effective for personalization, and why is it effective in those cases? To address the first question, we adapt a state-of-the-art automatic speech recognition (ASR) model to target speakers from four benchmark datasets representative of different speaker types. We show that ASR personalization with synthetic data is effective in all cases, but particularly when (i) the target speaker is underrepresented in the global data, and (ii) the capacity of the global model is limited. To address the second question of why personalized synthetic data is effective, we use controllable speech synthesis (CSS) to generate speech with varied styles and content. Surprisingly, we find that the text content of the synthetic data, rather than style, is important for speaker adaptation. These results lead us to propose a data selection strategy for ASR personalization based on speech content.

\end{abstract}
\section{Introduction}

\label{sec:intro}
Personalizing end-to-end automatic speech recognition (ASR) models is difficult due to the large number of parameters and limited speaker-specific data \cite{huang2020using}. Recent works propose to tackle the data scarcity using personalized text-to-speech synthesis (TTS) \cite{huang2020using, huang2020rapid,yue2021exploring}.  These works synthesize hours of personalized speech and combine it with small amounts of real speech to adapt generic ASR models to specific speakers. The success of this strategy is generally attributed to advancements in text-to-speech technologies, which enable high-quality speech synthesis in a target speaker’s voice.

\begin{figure}
    \centering
    \includegraphics[scale=0.27]{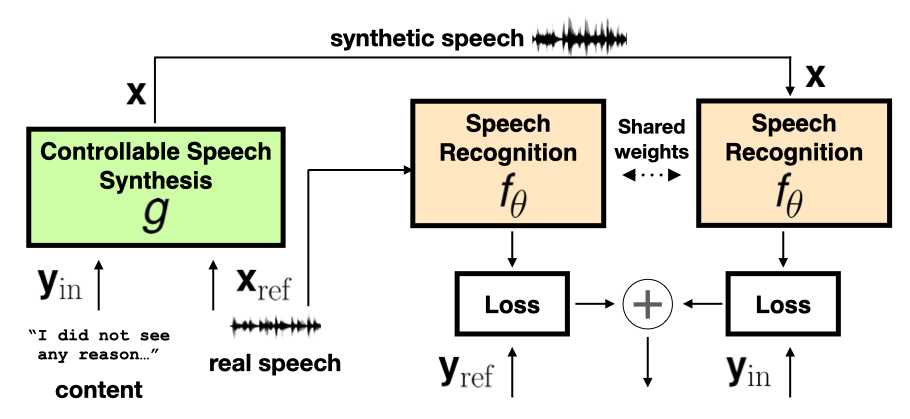}
    \caption{\textbf{Schematic of ASR personalization setup.} We use controllable speech synthesis (green) to generate personalized synthetic speech and adapt a speech recognition model (orange) to different types of speakers. See Section \ref{sec:method} for details.}
    \label{fig:main}
\end{figure}

In this paper, we perform comprehensive studies of this ASR personalization strategy across multiple datasets using a state-of-the-art controllable speech synthesis (CSS) model \cite{chang2022style}, as shown in Figure \ref{fig:main}. The CSS model allows us to vary the content and style of the personalized synthetic speech to understand their effects on model adaptation. Our most surprising insight is that matching the style of the synthetic speech to the target speaker is \emph{not} crucial for adaptation— ASR models adapt just as well with synthetic speech from other voices as they do with personalized synthetic speech. In contrast, matching the text content of the synthetic speech to that of the target speaker is important. These results lead us to rethink the conventional hypothesis to why personalized synthetic speech is helpful to ASR personalization, which we discuss in Section \ref{sec:why}. 

Overall, our experiments across multiple datasets provide novel insights about ASR personalization with synthetic data, addressing questions of when it works, why it works, and how to make it better:
\begin{itemize}[leftmargin=*,noitemsep,topsep=0pt]
\item Section \ref{sec:method} establishes our experimental setup of personalizing a state-of-the-art ASR model to four sets of target speakers using a state-of-the-art CSS model. 
\item \emph{When it works}— Section \ref{sec:when} highlights two scenarios that influence when ASR personalization with synthetic data is effective: (i) the target speaker is underrepresented in the global data, and (ii) the capacity of the model is limited.
\item \emph{Why it works}— Section \ref{sec:why} shows that the style of synthetic speech is not important for ASR personalization and poses alternative explanations for the utility of synthetic data in ASR personalization. 
\item \emph{How to make it better}— Section \ref{sec:how} evaluates the gradient matching strategy for improving the selection of synthetic data content for ASR personalization.
\end{itemize}

\section{Related Work}
Prior work addresses the data sparsity issue in ASR personalization by representing speaker dependence with condensed vectors \cite{saon2013speaker,snyder2018x,
fan2019speaker,sari2020unsupervised}, reducing the parameters that are updated during model adaptation \cite{
swietojanski2016learning, liao2013speaker,tobin2022personalized}, or with regularization techniques \cite{yu2013kl}. See \cite{bell2020adaptation} for a comprehensive review of speaker adaptation approaches for ASR.
%

More recently, some works have proposed adapting ASR models using data from personalized TTS. Huang \emph{et al.}~initially demonstrated this strategy for a hybrid ASR model \cite{huang2020using} and an RNN-T model \cite{huang2020rapid}. Later, Yue \emph{et al.}~used this strategy to adapt a LAS model trained on audiobook data to public speakers \cite{yue2021exploring}. 
In this paper, we build on this line of work by performing comprehensive ASR personalization experiments across multiple datasets using synthetic data from a state-of-the-art CSS model \cite{chang2022style}. 

Also related to our paper, existing works have used TTS models to augment the amount of training data for ASR models \cite{rossenbach2020generating, du2020speaker, rosenberg2019speech, mimura2018leveraging, hu2022synt++}, though these works are outside of the context of personalization. 

\section{Experimental Setup} \label{sec:method}

\subsection{Problem Formulation}
Let $\vec{x} \in \mathbb{R}^T$ denote a speech sample, and let $\vec{y} = [y_1, \cdots, y_C]$ denote the text content of $\vec{x}$. Let $f_\theta$ be a speech recognition model that maps from $\vec{x}$ to the %
conditional distribution $p(\vec{y} | \vec{x})$. Our goal is to train this model to work well for a specific target individual, i.e., we aim to minimize,
\begin{equation*}\label{eq:obj}
    \mathcal{L}(p_{\textrm{person}}; \theta) := \mathbb{E}_{(\vec{x},\vec{y}) \sim p_{\textrm{person}}(\vec{x}, \vec{y})}~ \ell(f_\theta(\vec{x}), \vec{y}),
\end{equation*}
\useshortskip where $p_{\textrm{person}}$ is the speech distribution of the target individual and $\ell$ is a text recognition loss (i.e., CTC loss \cite{graves2006connectionist}). However, we only have access to a very limited personalized dataset, $\mathrm{D}_{\textrm{person}} = \{(\vec{x}^{(i)}, \vec{y}^{(i)}) \sim p_{\textrm{person}}\}_{i=0}^M$, %
such that directly optimizing $f_\theta$ with respect to $\mathcal{L}(\mathrm{D}_{\textrm{person}}; \theta)$ 
results in overfitting.

\begin{table}[t]
    \setlength{\tabcolsep}{0.3em} 
    \newcommand{\indentrow}{\hspace{6pt}}
    
    \centering
    \footnotesize
    \resizebox{1\columnwidth}{!}{  
        \begin{tabular}{llc|cccc}
            \toprule
            && \textbf{Word Error Rate (\%)} & \textbf{Global Model} & \textbf{Pers. Model} & \textbf{Pers. Model} \\
            && \textbf{Adaptation Data} & N/A & $\textrm{D}_\textrm{person}$ & $\textrm{D}_\textrm{person} + \textrm{D}_\textrm{person-syn}$ \\
            \midrule
            \multirow{4}{*}{\rot{\textbf{Speaker}}} & \multirow{4}{*}{\rot{\textbf{Category}}} 
            & \textbf{1)} LS test-clean \cite{panayotov2015librispeech} & 3.5 & 3.8 (-8.6\% $\downarrow$) & \textbf{3.4} (2.9\% $\downarrow$) \\
            && \textbf{2)} LS test-other \cite{panayotov2015librispeech}& 9.1 & 8.5 (6.6\% $\downarrow$) & \textbf{7.8} (14.3\% $\downarrow$) \\
            && \textbf{3)} LJSpeech \cite{ljspeech17} & 7.7 & 6.4 (16.9\% $\downarrow$) & \textbf{3.2} (58.4\% $\downarrow$) \\
            && \textbf{4)} TED-LIUM 3 \cite{hernandez2018ted} & 11.3 & 10.8 (4.4\% $\downarrow$) & \textbf{9.5} (15.9\% $\downarrow$)\\
            \bottomrule   
        \end{tabular}
    }
    \caption{\textbf{ASR personalization results across different speaker categories.} Greater improvements are observed from speakers who are underrepresented in the global data. The metric shown is word error rate (WER); lower is better.}
    \label{table:personalization-results-main}
\end{table}

\subsection{ASR Personalization with Synthetic Data}

To tackle the data scarcity problem, we first optimize $f_\theta$ with respect to $\mathcal{L}(\mathrm{D}_{\textrm{global}}; \theta)$ to obtain a global model, where $\mathrm{D}_{\textrm{global}}$ is a very large, multi-speaker dataset.
Subsequently,
we finetune $f_\theta$ on a combination of real and synthetic data to obtain a personalized model.
We construct the personalized synthetic dataset $\mathrm{D}_{\textrm{person-syn}} = \{(\vec{x}^{(i)}, \vec{y}^{(i)})\}_{i=0}^N$ by sampling text from the target language domain $p_{\textrm{target-txt}}$ and passing it through a text-to-speech synthesis model $p_{\textrm{person-syn}}(\vec{x} | \vec{y})$:
\begin{equation}\label{eq:D-person-syn}
\quad \vec{x}^{(i)} \sim p_{\textrm{person-syn}}(\vec{x} | \vec{y}=\vec{y}^{(i)} ), \quad \vec{y}^{(i)} \sim p_{\textrm{target-txt}}(\vec{y}), 
\end{equation}
where $p_{\textrm{person-syn}}(\vec{x} | \vec{y})$ is meant to approximate the personalized conditional distribution $p_{\textrm{person}}(\vec{x} | \vec{y})$.
The weighted combination of losses used for fine-tuning is given by:
$$\frac{1}{2}\left[\mathcal{L}(\mathrm{D}_{\textrm{person}}; \theta) + \mathcal{L}(\mathrm{D}_{\textrm{person-syn}}; \theta)\right].$$

In practice, we take $f_\theta$ to be the ESPNet model \cite{watanabe2018espnet} consisting of a conformer \cite{gulati2020conformer} encoder and a transformer \cite{vaswani2017attention} decoder. We take the combined training splits of the LibriSpeech dataset \cite{panayotov2015librispeech} as $\mathrm{D}_{\textrm{global}}$, which we use to train the global model before fine-tuning.

\begin{figure}
    \centering
    \includegraphics[scale=0.6]{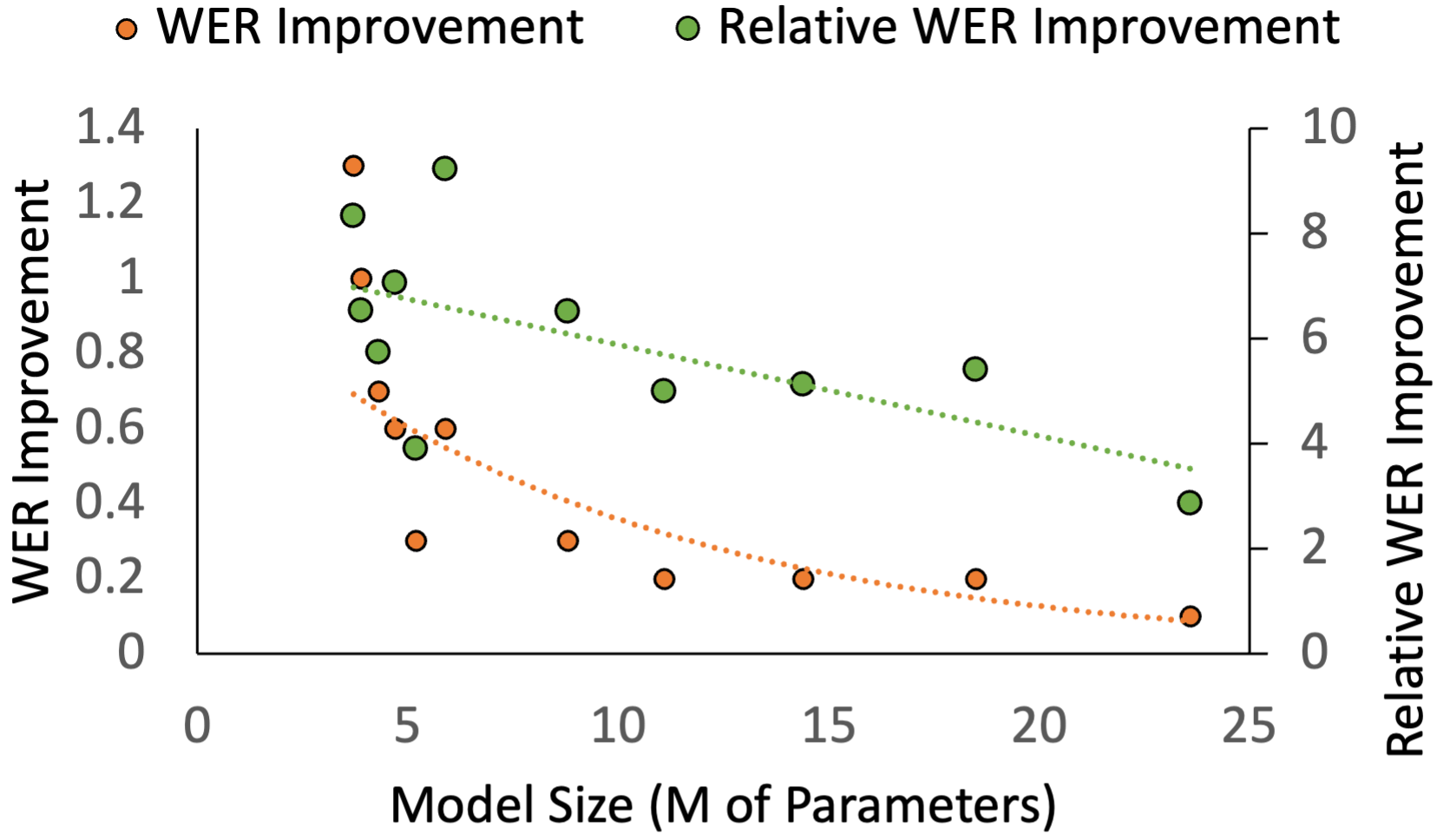}
    \vspace{-0.2cm}
    \caption{\textbf{ASR personalization results for different model sizes}. Personalization improves performance more for smaller models, both in terms of WER (orange) and relative WER (green). See text for details.}
    \label{fig:model-size}
\end{figure}

\begin{table*}[t]
    \setlength{\tabcolsep}{0.3em} 
    \newcommand{\indentrow}{\hspace{6pt}}
    
    \centering
    \footnotesize
    \resizebox{1.6\columnwidth}{!}{  
        \begin{tabular}{llccccc}
            \toprule
            & & & \multicolumn{4}{c}{\textbf{Speaker Category}} \\
            \cmidrule(lr){4-7}
            &\textbf{Model} & \textbf{Adaptation Data} & \textbf{1)} LS test-clean \cite{panayotov2015librispeech} & \textbf{2)} LS test-other \cite{panayotov2015librispeech} & \textbf{3)} LJSpeech \cite{ljspeech17} & \textbf{4)} TED-LIUM 3 \cite{hernandez2018ted}  \\
            \cmidrule(lr){2-2} \cmidrule(lr){3-3}\cmidrule(lr){4-4}\cmidrule(lr){5-5} \cmidrule(lr){6-6} \cmidrule(lr){7-7}
            \textbf{1} & Global & N/A & 8.5 & 17.5 & 7.7 & 11.3 \\
            \midrule
            \textbf{2} & Personalized\quad\quad\quad   & $\textrm{D}_\textrm{person}$ & 8.5 & 16.4 & 6.4 & 10.8 \\
            \textbf{3} & Personalized & $\textrm{D}_\textrm{person} + \textrm{D}_\textrm{person-syn}$ & 7.9 & 15.8 & 3.2 & 9.5 \\
            \midrule
            \textbf{4} & Personalized & $\textrm{D}_\textrm{person} + \textrm{D}_\textrm{other-person-syn}$ & 8.0 & 15.7 & 3.2 &  9.8 \\
            \textbf{5} & Personalized & $\textrm{D}_\textrm{person} + \textrm{D}_\textrm{multi-person-syn}$ & 7.7 & 16.1 & 3.9 & 9.2 \\
            \textbf{6} & Personalized & $\textrm{D}_\textrm{person} + \textrm{D}_\textrm{global-text-syn}$ & - & - & 6.0 & 10.9  \\
            \midrule
            \textbf{7} & Personalized & $\textrm{D}_\textrm{person} + \textrm{D}_\textrm{global}$ & 8.1 & 15.7 & 6.3 & 10.8 \\
            \bottomrule
            
        \end{tabular}
    }
    \caption{\textbf{ASR personalization results with different synthetic adaptation data}. Matching the synthetic data to the target speaker's content, but not style, is crucial to personalization. The metric shown is WER; lower is better. See text for details.}
    \label{table:personalization-results-ablation}
\end{table*}

\subsection{Controllable Speech Synthesis}\label{sec:css}

To perform personalized text-to-speech synthesis, we use a state-of-the-art controllable speech synthesis (CSS) model \cite{chang2022style} denoted by $g$. Given a reference speech sample $\vec{x}_\textrm{ref}$ and a text input $\vec{y}_\textrm{in}$, $g$ synthesizes speech containing the text $\vec{y}_\textrm{in}$ in the style of $\vec{x}_\textrm{ref}$, where style refers to all speech information aside from the text (i.e., speaker's voice, accent, acoustic environment, etc.). Since we have access to a real personalized dataset $\textrm{D}_{\textrm{person}}$, given an input text $\vec{y}_\textrm{in}$, we can generate samples from $p_{\textrm{person-syn}}(\vec{x} | \vec{y} = \vec{y}_\textrm{in})$ in Equation \ref{eq:D-person-syn} as follows:
\begin{align*}
\vec{x} \sim g(\vec{x}_\textrm{ref}, \vec{y}_\textrm{in}), \quad (\vec{x}_\textrm{ref}, \_) \sim \mathrm{D}_{\textrm{person}}.
\end{align*}
The full pipeline, in which synthetic data sampled from $g$ is used to finetune $f_\theta$, is shown in Figure \ref{fig:main}. In practice, we train $g$ on all training splits of the LibriTTS dataset \cite{zen2019libritts}.

\noindent \emph{Varying synthetic style and content.} The separate control of the style and content of $g$ enables us to experiment with different variations of synthetic data. 
Specifically, we also perform ASR personalization experiments with the following datasets in place of $\textrm{D}_{\textrm{person-syn}}$:
\begin{itemize}[leftmargin=*,noitemsep,topsep=0pt]
    \item \textbf{$\textrm{D}_{\textrm{other-person-syn}}$}: Instead of using samples from $\textrm{D}_{\textrm{person}}$ as style reference, we use samples from a different speaker's dataset: $(\vec{x}_\textrm{ref}, \_) \sim \mathrm{D}_{\textrm{other-person}}$.
    \item \textbf{$\textrm{D}_{\textrm{multi-person-syn}}$}: Instead of using samples from $\textrm{D}_{\textrm{person}}$ as style reference, we use samples from the global dataset: $(\vec{x}_\textrm{ref}, \_) \sim \mathrm{D}_{\textrm{global}}$.
    \item \textbf{$\textrm{D}_{\textrm{global-text-syn}}$}: Instead of sampling text from the target domain $p_{\textrm{target-txt}}$, we instead use text samples from the global dataset: $(\_, \vec{y}_\textrm{in}) \sim \mathrm{D}_{\textrm{global}}$.
\end{itemize}

The first two datasets enable us to evaluate the importance of matching the target speaker's style in the synthetic data. Note that these datasets could be produced by standard personalized TTS models rather than the CSS model, but this would require fine-tuning many TTS models. The third dataset enables us to evaluate the importance of matching the target speaker's language domain in the synthetic data.
\subsection{Personalization Datasets}
To understand when ASR personalization with synthetic data might be effective for different types of speakers, we perform our experiments on four sets of speakers representing different adaptation scenarios.

\begin{itemize}[leftmargin=*,noitemsep,topsep=0pt]
    \item \textbf{Category 1: The style of $p_\textrm{person}$ and the text of $p_{\textrm{target-txt}}$ are well-represented in $\textrm{D}_\textrm{global}$.} These are speakers who are considered canonical in the global data. 
    For this category, we use the speakers from the test-clean split of LibriSpeech \cite{panayotov2015librispeech}.
    \item \textbf{Category 2: The text domain of $p_{\textrm{target-txt}}$ matches $\textrm{D}_\textrm{global}$, but the style of $p_\textrm{person}$ is under-represented.} These are speakers whose styles may be considered less common in the global data. 
    For this category, we use the speakers from the test-other split of LibriSpeech \cite{panayotov2015librispeech}.
    \item \textbf{Category 3: The style of $p_\textrm{person}$ is well-represented in $\textrm{D}_\textrm{global}$, but the text domains do not match.} These are speakers whose styles are considered canonical in the global data, but who may be using different terms or phrases, e.g., the target speaker uses speech recognition for specific applications that differ from the settings where global data is collected. For this category, we use the speaker from LJSpeech \cite{ljspeech17}.
    \item \textbf{Category 4: The style of $p_\textrm{person}$ is under-represented in $\textrm{D}_\textrm{global}$ and the text domains do not match.} These are speakers who are not well-represented by $\textrm{D}_\textrm{global}$, combining both the challenges of Case 2 and Case 4. For this category, we use the speakers from TED-LIUM 3 \cite{hernandez2018ted}.
\end{itemize}
We take $\mathrm{D}_\textrm{person}$ to be a subsample of 1 minute from the speakers' original data. For the LJSpeech speaker, we use the validation and test splits from the original dataset. For the other speakers, we split the remainder of the data equally between validation and test sets.


\section{Results and Discussion}

\subsection{When does it work?}\label{sec:when}

\emph{Speaker category.} 
The quantitative results of our personalization experiments across four speaker categories are shown in Table \ref{table:personalization-results-main}. 
We find that finetuning using a combination of one minute of real personalized data and a large amount of synthetic personalized data (i.e., $\mathrm{D}_\textrm{person} + \mathrm{D}_\textrm{person-syn}$) improves performance across the board for all categories of speakers and outperforms using only real data (i.e., $\mathrm{D}_\textrm{person}$), which is not helpful for speakers from Category 1. 
Personalization with synthetic data is more effective when the \textbf{target speaker is underrepresented in the global data}; improvements are much larger for speakers from Categories 2-4 (relative WER improvement of 14.3-58\%) compared to speakers from Category 1 (relative WER improvement of 2.9\%).
This is consistent with the literature, where 
personalization is mainly used to adapt global models to target speakers that are out-of-domain \cite{green2021automatic,tobin2022personalized}. 

\noindent \emph{Model capacity.} How do we explain the modest improvement from personalization for speakers that are well-represented by the global data (i.e., Category 1 speakers)? One explanation lies in model capacity. While an ASR model with infinite capacity can, under certain assumptions, capture $p_\textrm{person}(\vec{y} | \vec{x})$ for all represented speakers in the global data, in practice, models only have limited capacity to capture each speaker's distribution. Personalization may improve performance by allowing a global model with limited capacity to specialize to one speaker in its domain. 

To test the hypothesis that model capacity influences personalization efficacy, 
we perform magnitude pruning and fine-tuning \cite{zhu2017prune} of the ASR model to reduce its capacity from 23.6M to 3.7M. We take model checkpoints at intermediate sizes and perform personalization to speakers that are well-represented in the global data (Category 1). The results are shown in Figure \ref{fig:model-size}. In support of our hypothesis, we find that \textbf{ASR models with smaller capacity benefit much more from personalization}, both in terms of WER improvement (1.3 for 3.7M vs. 0.1 for 23.6M) and relative WER improvement (8.3\% for 3.7M vs. 2.9\% for 23.6M).

\begin{table}[t]
    \setlength{\tabcolsep}{0.3em} 
    \newcommand{\indentrow}{\hspace{6pt}}
    
    \centering
    \footnotesize
    \resizebox{1.0\columnwidth}{!}{  
        \begin{tabular}{lcc}
            \toprule
            \textbf{Model} & Validation Set WER & Test Set WER \\
            \cmidrule(lr){1-1}\cmidrule(lr){2-2}\cmidrule(lr){3-3}
            \indentrow {Personalized Model} & 10.9 & 9.5 \\
            \indentrow {+ Gradient Matching} & \textbf{9.4} & \textbf{9.3} \\
        \end{tabular}
    }
    \caption{\textbf{Synthetic data selection using gradient matching improves personalization performance.} We compare default data selection to data selection with gradient matching for ASR personalization on TED-LIUM 3 speakers \cite{hernandez2018ted}. Metrics shown are WER; lower is better. See text for details.}
    \label{table:gradient-matching}
\end{table}

\subsection{Why does it work?}\label{sec:why}
Next, we ask why the personalized synthetic speech $\textrm{D}_\textrm{person-syn}$ benefits ASR personalization. The conventional hypothesis is that the style of the synthetic speech matches that of the target speaker; thus synthetic data augment the total amount of personalized data available to train the model. We test whether this is the case by changing the style and content for our CSS model and then adapting the ASR model on these alternate synthetic datasets as described in Section \ref{sec:css}. 

The quantitative evaluation is shown in Table \ref{table:personalization-results-ablation}. Surprisingly, we find that changing the style of the synthetic speech does not significantly degrade the results. For speakers with no language domain shift (i.e., Category 1-2), the results of personalizing with real data combined with personalized synthetic data (row 3) and synthetic data targeted to other speaker styles (row 4-5) are reasonably similar, with no clear winner. All of them outperform personalizing with only real data (row 2). The results suggest there is a benefit to using the synthetic data, but it is not because of the personalized speech style; rather, \textbf{synthetic speech appears to help regularize the model's language understanding}, since the large text corpus is the only element shared between these variations of synthetic data. In other words, synthetic speech plays a similar role to global data in data interpolation strategies for personalization \cite{mansour2020three}, in which global models are fine-tuned on a combination of real global data (for regularization) and real personalized data (for adaptation). We show results of data interpolation in row 7, which are similar to results of combining real and synthetic speech.

For those speakers with language domain gap (i.e., Category 3-4), we also observe that changing the style of the synthetic data does not significantly degrade results (rows 3-5). On the other hand, changing the text content from the target speaker's domain to the original global data domain (row 6) or replacing synthetic data with global data in data interpolation (row 7) significantly degrades the results and we no longer observe a large improvement over using only real data (row 2). These results indicate that even when the speaker style is mismatched, \textbf{synthetic speech provides the signal for adapting the ASR model to the target speaker's content}. 

\subsection{How to make it better?}\label{sec:how}

Since our investigation of the personalized synthetic speech revealed that its text content, rather than style, was important for ASR personalization, we investigate the efficacy of gradient matching \cite{killamsetty2021grad} for synthetic content selection. 
During fine-tuning, we periodically compute the gradient of the loss over the personalized validation set with respect to the current parameters, i.e., $G_\textrm{val} := \nabla_\theta \mathcal{L}(\mathrm{D}_{\textrm{person-val}}; \theta_t)$. Then, for each candidate synthetic utterance $\textrm{d} \in \mathrm{D}_\textrm{person-syn}$, we compute the gradient of the training loss with respect to the current parameters, i.e., 
$$G_\textrm{d} := \nabla_\theta \frac{1}{2}\left( \mathcal{L}(d; \theta_t) + \mathcal{L}(\mathrm{D}_\textrm{person}; \theta_t) \right)$$
Finally, we select the top synthetic utterances based on cosine similarity between $G_\textrm{val}$ and $G_\textrm{d}$ and train for a fixed number of epochs before re-selecting the synthetic data. 


We performed the experiments on the TEDLIUM 3 dataset. The results are shown in Table \ref{table:gradient-matching}. We found that using gradient matching yields a large gain on the validation set, which translates to a small gain on the test dataset. We draw two conclusions from these results: (i) selecting the correct content is helpful for improving performance, and (ii) a key direction for future work is to develop new data selection methods tailored for personalization, since it is easy for data selection methods to overfit on a small validation set.

\section{Conclusion}
Overall, our experiments across multiple datasets provide novel insights about ASR personalization with synthetic data, addressing questions of when it works, why it works, and how to make it better. One surprising finding is that style of personalized synthetic speech is not primarily responsible for the improvements we observe from using synthetic data for personalization-- the text is what is important, even if it is uttered in a different speaker's voice. 
We encourage future studies on this topic to adopt our controllable speech synthesis framework to disentangle the improvements that come from synthetic speech content vs. style.

\vfill\pagebreak

{\small
\bibliographystyle{IEEE}
\bibliography{refs}
}

\end{document}